%% file: main.tex
\documentclass[12pt,a4paper]{article}

\usepackage[cp1250]{inputenc}
\usepackage[round]{natbib} 
\usepackage{amsfonts,amsmath} 
\usepackage{amsthm}
\usepackage{enumerate} 
\usepackage{graphicx} 
\usepackage{authblk} 
\usepackage{epstopdf} 
\usepackage{floatrow}
\usepackage[top=1in, bottom=1in, left=1in, right=1in]{geometry}
\usepackage[font={small,it}]{caption}

\usepackage{textcomp}

\graphicspath{{img/}}

\usepackage{amssymb}
\usepackage{tikz}
\newcommand{\R}{\mathbb{R}}
\newcommand{\N}{\mathbb{N}}
\newcommand{\Z}{\mathbb{Z}}
\DeclareMathOperator*{\argmin}{argmin}
\DeclareMathOperator*{\argmax}{argmax}
\usepackage{bm}
\usepackage{siunitx}
\usepackage{setspace}
\usepackage{fancyhdr}
\usepackage{lipsum}

\title{\bf Description of ore particles from XMT images, supported by SEM-based image analysis}
\author{Orkun Furat$^{1}$, Thomas Lei{\ss}ner$^{2}$, Ralf Ditscherlein$^{2}$, Ond\v{r}ej \v{S}ediv\'{y}$^{1}$, Matthias Weber$^{1}$, Kai Bachmann$^{3}$, Jens Gutzmer$^{3}$, Urs Peuker$^{2}$, Volker~Schmidt$^{1}$}
\affil{\small \it $^1$Institute of Stochastics, Ulm University, D-89069 Ulm, Germany \\ 
\small \it $^2$Institute of Mechanical Process Engineering and Mineral Processing, Technische Universit\"{a}t Bergakademie Freiberg, D-09599 Freiberg, Germany \\ 
\small \it $^3$Helmholtz-Zentrum Dresden-Rossendorf, D-01328 Dresden, Germany, Helmholtz Institute Freiberg for Resource Technology, D-09599 Freiberg, Germany}
\date{}

\begin{document}
\maketitle
\sloppy 

\begin{abstract}
\noindent 
In this paper, 3D image data of ore particle systems is investigated. By combining X-ray micro tomography (XMT) with scanning electron microscope (SEM) based image analysis additional information about the mineralogical composition from certain planar sections can be gained. For the analysis of tomographic images of particle systems the extraction of single particles is essential. This is performed with a marker-based watershed algorithm and a post-processing step utilizing a neural network to reduce oversegmentation. The results are validated by comparing the 3D particle-wise segmentation empirically with 2D SEM images which have been obtained with a different imaging process and segmentation algorithm. Finally, a stereological application is shown, in which planar SEM images are embedded into the tomographic 3D image. This allows the estimation of local X-ray attenuation coefficients, which are material-specific quantities, in the entire tomographic image.
\end{abstract}

\section{Introduction}

The characterization of drill core sections, crushed rock and particles of different products from processing is essential for the mining and mineral processing industry. Methods for characterization are as numerous as the parameters to be determined. The subject of major interest in characterization is the 3D morphology of minerals and particles as well as the composition of particles and the spatial distribution of different minerals within the particles.
For a long time, morphological parameters such as particle size and chemical assays have been determined separately and set into correlation afterwards. Many well-accepted and precise analytical methods exist in this field such as sieve analysis, laser diffraction, chemical analysis and X-ray fluorescence spectroscopy.

Since the late 1980s, combined methods for simultaneous acquisition of morphological and compositional data have been available and they have seen a substantial development. In recent years, combined scanning electron microscopy (SEM) and energy dispersive X-ray spectroscopy (EDS) \citep{Sunderland1991} has become a standard method for the characterization of ores. It is performed on sliced and polished specimens, which can be a section of a drill core or an epoxy block containing the particles to be analyzed. 
Due to the two-dimensional nature of SEM-EDS, the characterization of three-dimensional features like size and volumetric composition exhibits a stereological bias.

On the other hand, using direct three-dimensional analysis, such as computed X-ray micro tomography (XMT), the problem of stereology disappears. In XMT of standard laboratory architecture, a specimen is penetrated by X-ray radiation at different angles of rotation. The acquired projection images are then reconstructed to a 3D volume representation of stacked 2D grayscale images. The grayscale value of a voxel (3D pixel) represents the X-ray attenuation coefficient of this volume element, which is a function of average atomic number, wavelength, thickness and density \citep{Gordzins1983}. 
However, this grayscale information does not always suffice for determining the mineralogical composition of the sample. Thus, XMT provides information about the 3D morphology of the specimen, but does not have the same characterization property which SEM-EDS provides for planar sections.

A correlative approach that combines both SEM-EDS and XMT has been utilized recently by Reyes and coworkers \citep{Reyes2017}. They compared SEM-EDX data with the corresponding section of registered volumetric data (2D XMT) gained by 3D XMT. In particular, they reported on 10 percent misclassified pyrite grains due to the segmentation based on a global thresholding algorithm. Furthermore, they reported on challenges arising from the different resolutions of the methods and the impact of partial volume artifacts.

In order to overcome such issues, sophisticated algorithms for image processing and analysis are essential for the characterization of particulate systems of multi-component materials such as ores. This involves the determination of different phases as well as particle-wise segmentation of volumetric data \citep{Cnudde2013,Maire2014,schlueter2014}.
Denoising is a common first step of image preprocessing. Linear filters like the Gaussian kernel \citep{Burger2010} are computationally feasible but have the disadvantage of blurring edges. Therefore, non-linear denoising methods like non-local means \citep{buades2005} can smooth images in homogeneous areas while preserving edges. Another important step for analyzing image data of particulate systems is the particle-wise segmentation such that each individual particle can be extracted for further analysis. A rather popular tool for segmentation is the watershed algorithm \citep{roerdink2001,soille2003}, which is a kind of region growth algorithm that operates on gradient images or distance maps of binarized images \citep{Burger2010}. A common issue of the watershed algorithm is that single particles are often divided into many segments, which is referred to as oversegmentation. Marker-based watershed algorithms \citep{spettl2015} can overcome this issue by a preceding determination of unique markers for each particle. Nevertheless, irregularly shaped particles, like non-spherical or elongated shapes, make the finding of unique markers difficult, such that post-precessing of segmentations achieved by the watershed algorithm may be necessary.
\\
\\
In this paper we present results of a correlation analysis of 2D SEM-EDS with 3D XMT images for a greisen-type ore. Therefore, volumetric image data of a sample was obtained via XMT, and for some planar sections of the same sample, 2D SEM-EDS data was obtained with the mineral liberation analyzer (MLA) scanning electron microscope and segmented by software provided by the manufacturer of the microscope. In order to combine these two imaging processes, we will describe a method for embedding (registering) the planar sections obtained via SEM-EDS in the volumetric data. Furthermore, we present a method for the particle-wise segmentation of the volumetric image data with a marker-based watershed algorithm and a post processing step to reduce oversegmentation. For that purpose, we trained a neural network to decide, based on local geometrical and grayscale features of the volumetric image, whether adjacent segments determined by the watershed algorithm should be merged. This method was validated by comparing particle size and particle shape distributions of the segmentation obtained by the MLA software and a corresponding 2D section from the volumetric segmentation method proposed in the present paper.

Since the 2D SEM-EDS images do not only contain information about the morphology of particles across the corresponding planar section, but also provide information about the mineralogical composition of the particles, we show how this additional information can be used to extrapolate the mineral classification from 2D SEM-EDS to the 3D XMT data by estimating local X-ray attenuation coefficients.

\section{Materials and methods}

\subsection{Material and sample preparation}
The material considered in this paper is a greisen-type ore from the Zinnwald/Cinovec deposit at the German-Czech border in the ``Erzgebirge''. The ore mainly comprises of quartz, topaz, zinnwaldite, muscovite and kaolinite. Mica compositions are twofold---including a mica of the siderophyllite - polylithionite series  (called ``zinnwaldite'' from hereon) and muscovite \citep{Rieder1998}. A large bulk sample was crushed and milled down to a particle size $<$ \SI{1}{\mm}. A size fraction of \SIrange{315}{500}{\micro\metre} was prepared by analytical sieving. Representative subsamples of this fraction, generated by a rotary sample divider, have been used for sample preparation and analysis as described below.

In the next step, an epoxy block was prepared. Therefore, the aliquot of \SI{2}{\gram} sample material mixed with  \SI{1}{\gram} graphite and \SI{2}{\gram} epoxy resin was used to prepare the grain mount. The resulting epoxy block was cut vertically in order to get sections in the direction of sedimentation. These sections were then tilted by $90$ degree and mounted again as \SI{20}{\mm} blocks (B-sections, see \citep{Heinig2015}). This sample underwent XMT measurement followed by grinding, polishing and analysis at the MLA.

\subsection{Computer tomography and mineral liberation analysis}
The grain mount was scanned using a Zeiss Xradia 510 Versa X-ray microscope. In order to attain volumetric data of a sufficient resolution as well as an appropriate size (3D field of view), a voxel size of $4.5$ \textmu m was chosen for the tomography. The parameters of the XMT scan are listed in Table~\ref{tab:ParameterOfXRM}.

\begin{table}
	\centering
    \caption{Parameters for the XMT scan of the considered sample.}
	\label{tab:ParameterOfXRM}
	\input{table1.tex}
\end{table}

The 3D volume reconstruction was done using the Zeiss XRM reconstructor software. This software works with a filtered back projection algorithm and an additional beam hardening correction method. A manual byte scaling was used to adjust the grayscale values of the histogram to the range of interest. The parameters used for volume reconstruction are listed in Table~\ref{tab:ParameterOfReconstruction}. After the volumetric XMT scan, additional SEM-EDS measurements of the sample were made with the MLA.

\begin{table}
	\centering
    \caption{Parameters for the reconstruction of the considered sample.}
	\label{tab:ParameterOfReconstruction}
	\input{table2.tex}
\end{table}

The system used for the additional MLA measurements consists of an FEI Quanta 650F scanning electron microscope equipped with two Bruker Quantax X-Flash 5030 energy-dispersive X-ray spectrometers and the MLA software suite, version 3.1.4, for automated data acquisition. In order to correlatively combine SEM-EDS with XMT the same sample that was used in the previous XMT measurements was grinded and polished multiple times for measurements at different planar sections with the MLA. Therefore, consistent operating conditions were applied for each considered planar section using the GXMAP measurement mode at \SI{20}{\kilo\volt} acceleration voltages, \SI{10}{\nano\ampere} probe current, \SI{1}{\micro\metre}/pixel, \SI{6}{\milli\second} acquisition time and a step size of $6$ pixels. 
The measurements at the MLA are automatically processed with its provided software, resulting in segmented false color 2D images of planar sections of the sample, where the colors in the false color image represent different minerals each, as seen in Figure~\ref{fig:registration} a). In total two such images were acquired at spatially different planar sections of the sample. From hereon we will refer to these false color images obtained by the MLA system as MLA images/data. More detailed information about the functionality of the MLA system can be found in \cite{Fandrich2007}, whereas details regarding data processing are shown in \cite{Bachmann2017}. Additional information on the measurements with the MLA of the ore used in this study can be found in \cite{Heinig2015}.

\subsection{Image processing}\label{sec:imageProcessing}

The selection of image processing tools and their careful adjustment to the experimental dataset has a critical effect on the subsequent analysis. In this section, all the additional image processing steps that were applied to the reconstructed 3D XMT image are explained in detail. We assume that the image is observed on a finite set of voxels $W^\prime \subset W$ being a subset of a bounded region $W\subset \mathbb{R}^3$. The 16-bit grayscale image obtained by the XMT measurement can be represented by a mapping $I\colon W^\prime \to \{0,\ldots,65535\}$ which assigns to each voxel $x\in W^\prime$ its grayscale value $I(x)$.

\subsubsection{Denoising and enhancement of edges}\label{sec:denoising}

A first image processing step is to reduce noise in the image. Here, we use the results of a detailed comparison of different algorithms described in~\cite{schlueter2014}, where the best quality denoising is achieved by a subsequent application of two operations -- non-local means and unsharp mask. 

\begin{figure}[!htb]
\begin{center}
\includegraphics[width=0.3\textwidth]{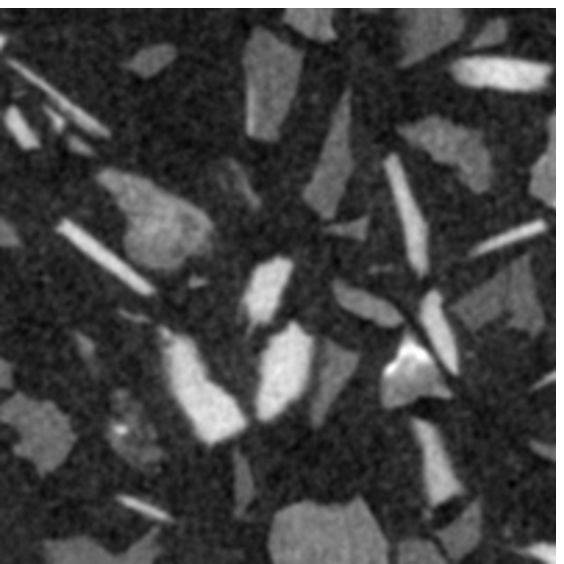} 
\includegraphics[width=0.3\textwidth]{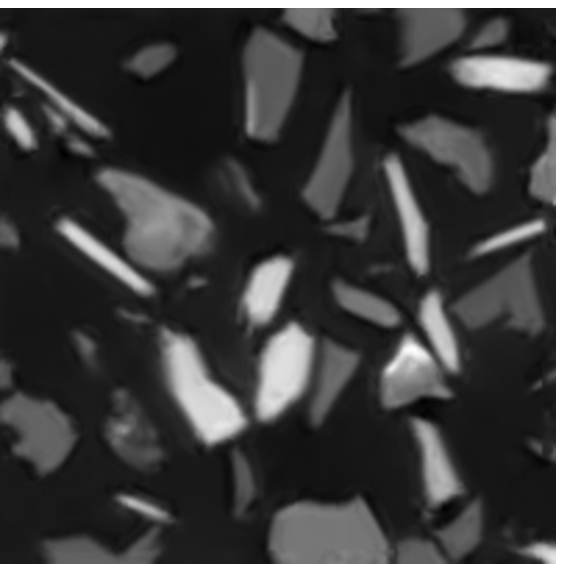} 
\includegraphics[width=0.3\textwidth]{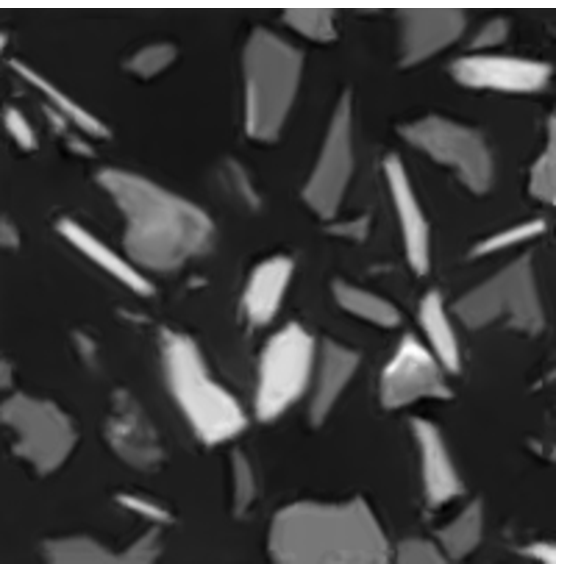} \\
\begin{minipage}{0.3\textwidth}{\centerline{a)}}\end{minipage}
\begin{minipage}{0.3\textwidth}{\centerline{b)}}\end{minipage}
\begin{minipage}{0.3\textwidth}{\centerline{c)}}\end{minipage} \vspace{3mm}\\ 
\includegraphics[width=0.3\textwidth]{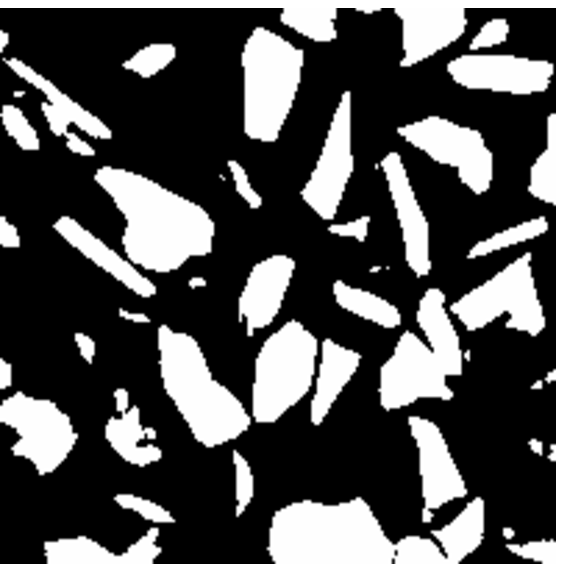} 
\includegraphics[width=0.3\textwidth]{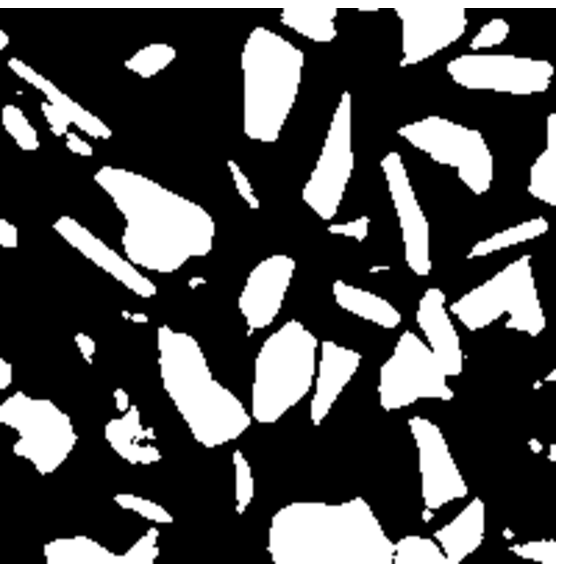} 
\includegraphics[width=0.3\textwidth]{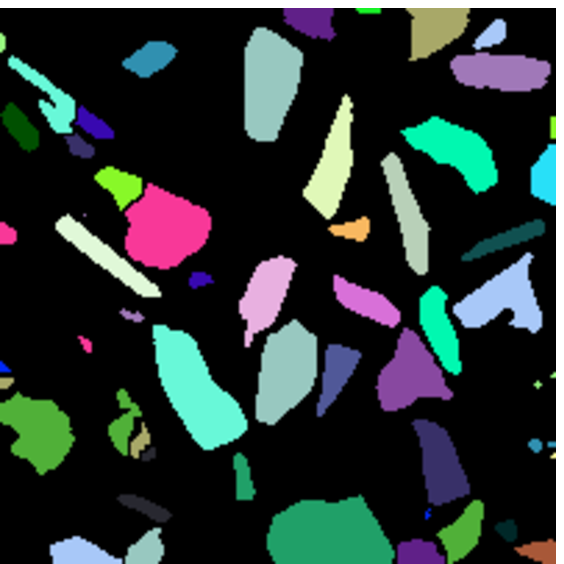} \\
\begin{minipage}{0.3\textwidth}{\centerline{d)}}\end{minipage}
\begin{minipage}{0.3\textwidth}{\centerline{e)}}\end{minipage}
\begin{minipage}{0.3\textwidth}{\centerline{f)}}\end{minipage}\\
\end{center}
\caption{Effect of image processing steps on a cutout of one slice of the sample. a) original grayscale image, b) image denoised by non-local means, c) image processed by unsharp masking, d) image binarized by local adaptive thresholding, e) image processed by an opening with a ball with a radius of one voxel, f) image segmented into particles by a marker-based watershed algorithm exhibiting some oversegmentation.}
\label{fig:processing}
\end{figure}

In the non-local means method \citep{buades2005}, the grayscale value of each voxel $x\in W^\prime$ is modified by some weighting function $w(x,y)$ applied to every voxel $y\in W^\prime$. The resulting image is obtained by
\begin{equation}\label{eq:NL}
I_{\rm{NL}}(x) = \sum_{y\in W^{\prime}} w(x,y) I(y).
\end{equation}
The weighting function suggested in \cite{buades2005} has the form
\begin{equation}\label{eq:NLweights}
w(x,y)=\frac{1}{Z(x)}\exp{\left(-\frac{1}{h^2}\sum_{z\in N} G_\sigma(z) |I(x+z)-I(y+z)|^2 \right)},
\end{equation}
where $G_\sigma$ is a Gaussian kernel with standard deviation $\sigma$, $Z(x)$ is a normalizing factor, $N$ is a local window centered at the origin $o\in \R^3$, and $h>0$ is a parameter influencing the level of filtering.
In (\ref{eq:NL}) and (\ref{eq:NLweights}) it can be seen that the value of the denoised image $I_{\rm{NL}}(x)$ at some voxel $x\in W^\prime$ is a weighted sum of grayscale values of the noisy image $I$ at some other positions $y\in W^\prime$. The weights $w(x,y)$ are large, when the voxels $x$ and $y$ have similar grayscale values in their neighborhood, where the size of the neighborhood is determined by the standard deviation $\sigma>0$ of the Gaussian kernel and the local window $N$.  However, if the positions $x+z, y+z$ lie outside of the observation window $W^\prime$ for some  $z\in N$ the corresponding summands are ignored in (\ref{eq:NLweights}).
Thus, the sum for the computation of the kernel in (\ref{eq:NLweights}) is restricted to a predefined window $N$, in order to keep the operation computationally feasible. 
The effect of non-local means denoising is shown in Figure~\ref{fig:processing}b).

The unsharp mask filter \citep{pratt2007} serves for the enhancement of edges in the image (surfaces in the 3D image). For the execution of this filter, a lower-resolution image $I_L$ is computed first, which is obtained by blurring the original image with a smoothing kernel. Then, the result is obtained as a weighted difference between the original and the lower-resolution image, i.e.
\begin{equation}
I_{\rm{UM}}({\bf x}) = \frac{c}{2c-1}I({\bf x}) + \frac{1-c}{2c-1}I_L ({\bf x}),
\end{equation}
where $c$ is a weighting constant which takes values typically in the range from $3/5$ to $5/6$. A 2D slice of the image processed by unsharp masking is shown in Figure~\ref{fig:processing}c).

\subsubsection{Binarization}\label{sec:binarization}

After having performed all the preliminary image processing operations described in Section~\ref{sec:denoising}, the grayscale image is binarized which means that the foreground phase (particles) is separated from the background. It is quite typical for XMT scans that the grayscale values are not globally consistent, i.e., the transitions between foreground and background are identified on different grayscale levels. This is often related to ring artifacts \citep{barrett2004} caused by a miscalibrated or defective detector element. They appear in 2D slices as concentric rings with different spanning of grayscale values. 

In order to avoid inconsistencies in the binarization, we apply a local adaptive thresholding technique using Sauvola's thresholds, as described in~\cite{shafait2008}. For each voxel $x\in W^\prime$, the local threshold $t(x)$ is determined by
\begin{equation}
t(x)=m(x)\left(1+k\left(\frac{s(x)}{R}-1 \right)\right),
\end{equation}
where $m( x)$ and $s(x)$ are the mean and standard deviation of the grayscale values in a local cuboidal window of the image $I$ centered at the voxel $x$, respectively.
The value of $R$ is the maximum value of the standard deviation achievable for given image type ($R=32768$ for 16-bit grayscale image) and $k$ is a parameter which regulates the height of the threshold and typically lies in $[0.2, 0.5]$, see~\cite{shafait2008}. 

The binarization $B\colon W^\prime \to \{0,1\}$ of a 16-bit grayscale image $I$ is given by
\begin{equation}
 B(x)=\left\{\begin{array}{ll} 1, & I(x)\geq t(x)\\
         0, & I(x) <t(x)\end{array}\right. .
\end{equation}
A cutout of a 2D slice  of the binarized 3D image is shown in Figure~\ref{fig:processing}d).

\subsubsection{Morphological smoothing}

Application of suitable morphological operations to the binarized data helps to avoid small artifacts and correct irregularities at the interfaces between foreground and background. We use a popular smoothing operation called opening, which is a composition of erosion and dilation. In mathematical terms, opening of a set $A_1$ by a set $A_2$ can be written as
\begin{equation}
A_1 \circ A_2=(A_1\ominus A_2)\oplus A_2,
\end{equation}
where $\oplus$ is the Minkowski addition (dilation) and $\ominus$ is the Minkowski difference (erosion), see~\cite{chiu2013}. In our setting, $A_1$ is the foreground phase of particles, i.e. the set of voxels $x\in W^\prime$ with $B(x)=1$, and $A_2$ is a ball with fixed radius of one or a few voxels. The effect of morphological smoothing is shown in Figure~\ref{fig:processing}e).
\subsection{Segmentation}\label{sec:segmentation}

Segmentation of particles is performed by a watershed algorithm \citep{roerdink2001}. In particular, we use a marker-based watershed transformation considering extended regional minima which has been described in detail in \cite{spettl2015}. This method determines markers based on the inverted Euclidean distance transform of the binarized image. However, since the particles have irregular shapes, problems can arise when applying the marker-based watershed. Especially elongated particles can lead to oversegmentation, since local minima of the inverted Euclidean distance transform of such particles extend over large areas. This makes it difficult to determine unique markers for these particles, which leads to oversegmentation, see Figure~\ref{fig:processing} f). We overcome this problem by using a neural network, see~\cite{hastie2009}, to determine whether two adjacent regions of an oversegmented image should be merged or not. Therefore, we give an introduction into simple feed-forward neural networks, which are used in our application.

\subsubsection{Neural networks}\label{subsec:NeuralNet}
\begin{figure}
\centering
\includegraphics{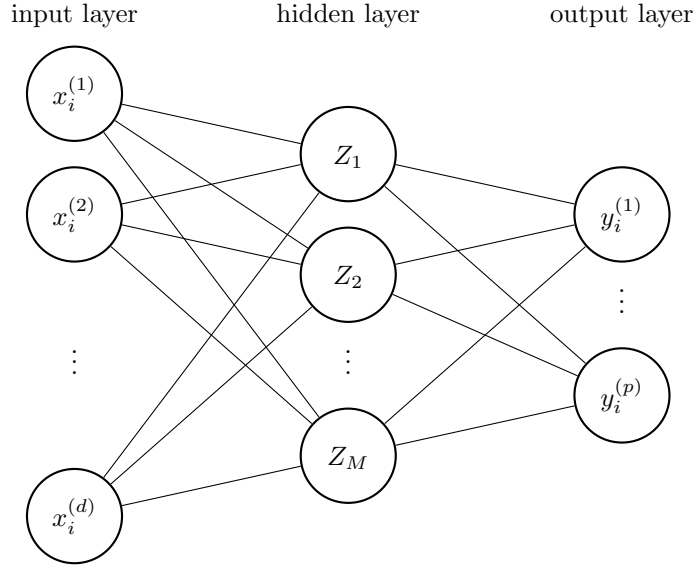}
\caption{Graph representation of a feed-forward network $\R^d\to \R^p$ with a single hidden layer with $M$ units.}\label{fig:neural}
\end{figure}

Neural networks are nonlinear regression models, which are often represented by a network diagram, see Figure \ref{fig:neural}. A typical neural network has one or several hidden layers, containing multiple units, so-called neurons, which process the input sequentially towards the output layer, yet more complex architectures are possible. For simplicity we describe such a feed-forward network with a single hidden layer.

Let $d, p, N \in \N$ be arbitrary positive integers. For a regression problem $f(x_i)=y_i$ with predictor variables $x_i=(x_i^{(1)},\dots, x_i^{(d)}) \in \R^d$ and response variables $y_i=(y_i^{(1)},\dots, y_i^{(p)}) \in \R^p$ for $i=1,\dots,N$, a typical regression model with a single layer network containing $M\in \N$ hidden units has the following representation~\citep{hastie2009}.
For an input (or feature) vector $x\in \R^d$ the hidden units $Z_1, \ldots, Z_M$ have the output
\begin{equation}\label{eq:hiddenLayer}
Z_m(x)=s{\left(\alpha_{0m}+\alpha_m^\top x\right)}, \: \text{for each } m=1,\dots, M,
\end{equation}
where $\alpha_{0m} \in \R$ and $\alpha_m \in \R^d$ are regression parameters and $s\colon \R \to \R$ is the so-called activation function. A typical choice for the activation function is $s =\tanh$.
The outputs of the hidden units, described by (\ref{eq:hiddenLayer}), are then passed to the output layer, which contains $p$ output units. This number is determined by the dimension of the response variables. The values of the output units $T_1, \ldots, T_p$ which are processed towards the output are given by
\begin{equation}\label{eq:outputLayer}
T_k(x)=\beta_{0k}+ \left(Z_1(x),\dots,Z_M(x)\right)^\top\!\beta_k,  \text{\: for each } k=1,\dots, p,
\end{equation}
where $\beta_{0k}\in \R$ and $\beta_k\in \R^M$ are regression parameters of the output units.
The final output $f(x)=(f_1(x),\dots,f_p(x))^\top$ of the neural network is given by evaluating certain output functions $g_1,\dots,g_p \colon \R^p \to \R$ with the vector $T(x)=\left(T_1(x),\dots,T_p(x)\right)$, i.e.,
\begin{equation}\label{eq:output}
f_j(x)=g_j(T(x)), \text{\: for each } j=1,\dots, p.
\end{equation}
The choice of output functions depends on the problem one is trying to solve. For simple regression problems linear output functions can be chosen, i.e., $g_k(T)=T_k$, whereas other problems which require the outputs of the neural network  to be normalized use output functions like $g_k(T)=\exp{\left(T_k\right)} \left(\sum_{l=1}^p \exp{\left(T_l\right)}\right)^{-1} $.

The described regression model, which is defined by (\ref{eq:hiddenLayer})--(\ref{eq:output})  has a large parameter space $\Theta$, and a network with a specific parameter constellation $\theta=(\alpha_{01},\dots,\alpha_{0M},\alpha_{1},\dots,\alpha_{M},\beta_{01},\dots,\beta_{0p},\beta_1,\dots,\beta_p)\in \Theta$ can be denoted by $f_\theta$. Training a neural network means finding an optimal regression parameter $\hat{\theta}\in \Theta$ such that $f_{\hat{\theta}}(x_i)\approx y_i$ for each $i=1,\dots,N$. For a model with response dimension $p=1$, this can be formulated as an optimization problem by using, for example, the sum of squared errors
\begin{equation}\label{eq:SSE}
\hat{\theta}=\argmin_{\theta\in\Theta} \sum_{i=1}^N \left(y_i-f_\theta(x_i)\right)^2,
\end{equation}
or the cross entropy
\begin{equation}\label{eq:CE}
\hat{\theta}=\argmin_{\theta\in\Theta} {\left(-\sum_{i=1}^N y_i \log{\left(f_\theta(x_i)\right)}+ (1-y_i) \log{\left(1-f_\theta(x_i)\right)}\right)}.
\end{equation}
In our application, the task of the neural network is to decide whether adjacent regions of an oversegmented image should be merged. Therefore, we use the cross entropy which is better suited for classification problems.
Since the error functions in (\ref{eq:SSE}) and (\ref{eq:CE}) are, due to the smoothness of the chosen activation and output functions, differentiable, the optimization is usually performed by some sort of gradient descent. Due to the large number of regression parameters some techniques, like validation and regularization, are used to avoid overfitting, see \cite{hastie2009}.

\subsubsection{Elimination of oversegmentation}

\begin{figure}[!htb]
\begin{center}
\includegraphics[width=0.45\textwidth]{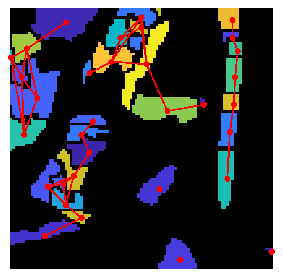} 
\includegraphics[width=0.45\textwidth]{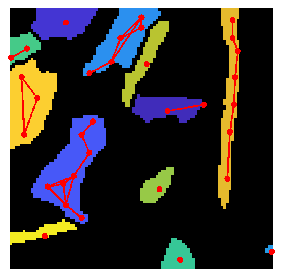} 
\begin{minipage}{0.45\textwidth}{\centerline{a)}}\end{minipage}
\begin{minipage}{0.45\textwidth}{\centerline{b)}}\end{minipage}
\end{center}
\caption{Application of a neural network to an oversegmented watershed image. a) Oversegmented watershed image with graph representation. Each red point represents a region, edges (red lines) are set between adjacent regions. b) Correct segmentation is achieved by removing edges between regions which should not be merged with the help of a neural network. The connected components of the resulting graph represent single regions.}
\label{fig:NeuralNetWatershed}
\end{figure}
\begin{figure}
\begin{center}
\includegraphics[width=0.6\textwidth]{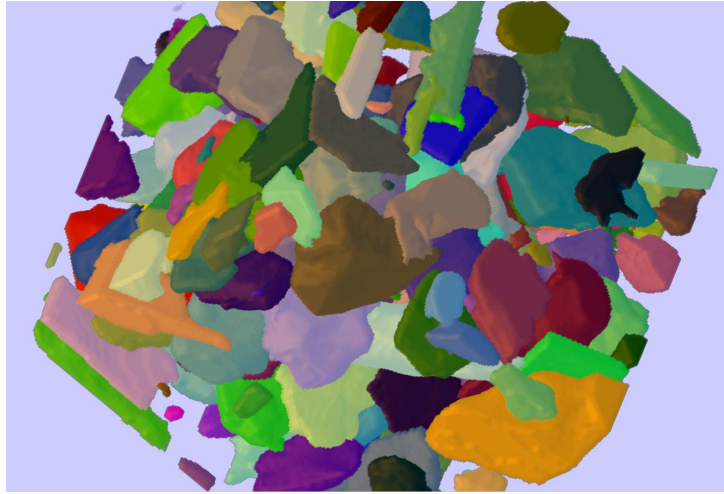} 
\caption{Cut-out of the volumetric particle-wise segmentation obtained by a marker-based watershed algorithm with a post-processing step which utilizes a neural network.}
\label{fig:volumetricSegmentation}
\end{center}
\end{figure}

In order to apply neural networks to an oversegmented image $I_{\text{over}}$ obtained by the marker-based watershed algorithm, we represent such an image as a graph $G=(V,E)$, where each vertex $v\in V$ represents a region of $I_{\text{over}}$. For each pair of adjacent regions $v_1,v_2 \in V$ in $I_{\text{over}}$, we set an edge $e=(v_1,v_2) \in E$ in the graph representation, see Figure~\ref{fig:NeuralNetWatershed}a). In order to receive an improved segmentation $I$ from the oversegmented image $I_{\text{over}}$, we have to remove edges between adjacent regions which belong to different particles. Then the connected components of this reduced graph $\hat{G}=(V,\hat{E})$, with $\hat{E}\subset E$,  represent the particles of the improved segmentation, i.e., a remaining edge $e=(v_1,v_2)\in \hat{E}$ tells us to merge the regions of the oversegmented image represented by $v_1$ and $v_2$, see Figure~\ref{fig:NeuralNetWatershed}b). More specifically, we have to find an edge function $w\colon E \to \{0,1\}$, where $w{\left((v_1,v_2)\right)}=1$ if and only if the two regions $v_1,v_2\in V$ should be merged. To begin with, we determine this function $w$ manually, based on a small oversegmented cut-out with the size of $200\times 200\times 200$ voxels, obtained by the marker-based watershed algorithm. So, for each edge $e\in E$, the value $w(e)\in \{0,1\}$ represents a response variable for our regression model. Now we have to find suitable features, which describe two adjacent regions of the watershed image in such a way that we are able to reliably decide whether they should be merged or not. Some features considered for two adjacent regions $v_1$ and $v_2$ are listed below:
\begin{enumerate}
\item The distribution of the grayscale values of the image around the watershed line between $v_1$ and $v_2$ and its first four moments.
\item The distribution and the first four moments of the absolute gradient values around the watershed line are considered. We calculated the absolute gradient of the grayscale image with Sobel operators, see \cite{soille2003}.
\end{enumerate}
The previously mentioned features solely consider local contrast information of the grayscale or gradient image. In order to geometrically describe the watershed line between $v_1$ and $v_2$, the following geometrical features are considered:
\begin{enumerate}
 \setcounter{enumi}{3}
\item The eigenvalues of the principal component analysis of the watershed line voxels, see \cite{hastie2009}.
\item The local curvature is another geometrical feature which is considered by the neural network.
\end{enumerate}
These feature vectors are calculated for each pair of adjacent regions $v_1$ and $v_2$. We denote them by $x_e \in \R^d$, where $e=(v_1,v_2)$ is the corresponding edge.
Now we can formulate our regression problem by $f(x_e)=w(e)$ for each edge $e\in E$.
As for the regression model, we have chosen a single layer network with $75$ hidden units and activation function $s=\tanh$ as described in Section \ref{subsec:NeuralNet}. The optimal number of hidden units has been determined by a grid-search algorithm, meaning that we trained the network for multiple hidden layer sizes and chose the configuration with the best performance. Since our response variables are one-dimensional, we have only one unit in the output layer for which we chose the output function $g(T)=1/(1+\exp{(-T)})$, which ensures that the output of the neural network belongs to $(0,1)$. After training the regression parameters of the neural network based on the manually segmented cut-out, the neural network, denoted by $f$, can be applied on new image data in the following way:
\begin{enumerate}
\item Compute an (oversegmented) image using the watershed algorithm and determine its graph representation $G=(V,E)$ where $E$ contains an edge for each pair of adjacent regions.
\item Determine the local features $x_e$ for each pair of adjacent regions.
Together with the neural network $f$, we receive a weight function $w\colon E\to (0,1)$ for the graph $G$, where $w(e)=f(x_e)$.
\item Reduce the weighted graph $(V,E,w)$. This is done by thresholding, i.e., $\hat{E}=\{e\in E : w(e)=f(x_e)\geq \lambda\}$ with a threshold $\lambda \in (0,1)$. Alternatively, graph clustering methods, see \cite{SCHAEFFER200727}, can be used to reduce the graph.
\item Determine the connected components of the reduced graph $\hat{G}=(V,\hat{E}).$ Merge regions of the same connected component in the oversegmented image.
\end{enumerate}
Figure \ref{fig:volumetricSegmentation} visualizes the result of the particle-wise segmentation of our proposed method.
\section{Results and discussion}
\subsection{Comparative analysis of 2D and 3D data}
In the previous section we described how the tomographic 3D data was segmented. Additionally we have two 2D MLA images from the same sample at a $10$ times higher resolution. The MLA data, which provides additional information about the mineralogical composition of particles, was segmented by the MLA software using a particle de-agglomeration algorithm, see \cite{Fandrich2007}.
Therefore we analyze the consistency of our approach to segment the 3D data and the segmented 2D MLA data. For that purpose we compare the distributions of particle sizes and shape characteristics of the MLA data with the corresponding distributions of planar sections of the segmented 3D particle system.

\subsubsection{Size characteristics}

Regarding the size characteristics for planar cross-section of a particle $P \subset \R^2$, we consider the area $a(P)$, the length of the perimeter $l(P)$, and the mean width $w(P)$.
These characteristics are closely related to intrinsic volumes, being basic descriptors of compact sets \citep{ohser2000}. 

While the area of a particle cross-section can be simply estimated by rescaling the number of voxels belonging to that particle, estimation of the perimeter requires a better approximation than direct computation of transitions between voxels. We use the cornercount estimator described in~\cite{klette2004}, where the contribution of each boundary voxel to the total boundary length of a given particle cross-section is given by a specific weight depending on its neighborhood.

The mean width, see \cite{ohser2000}, is defined by 
\begin{equation}
w(P)=\frac{1}{\pi}\int_{0}^{\pi} \left(\max{T_x{\left(M_\alpha P\right)}}- \min{T_x{\left( M_\alpha P\right)}}\right) \text{d} \alpha,
\end{equation}
where $T_x$ is the projection on the $x$-axis, i.e., $T_x{\left( (x_1,x_2) \right)}=x_1$, and $M_\alpha \in SO_2$ is a  matrix which describes a 2D rotation with an angle $\alpha$ and $M_\alpha P= \{M_\alpha x : x\in P \}$.  
\begin{figure}
\begin{center}
\includegraphics[width=0.3\textwidth]{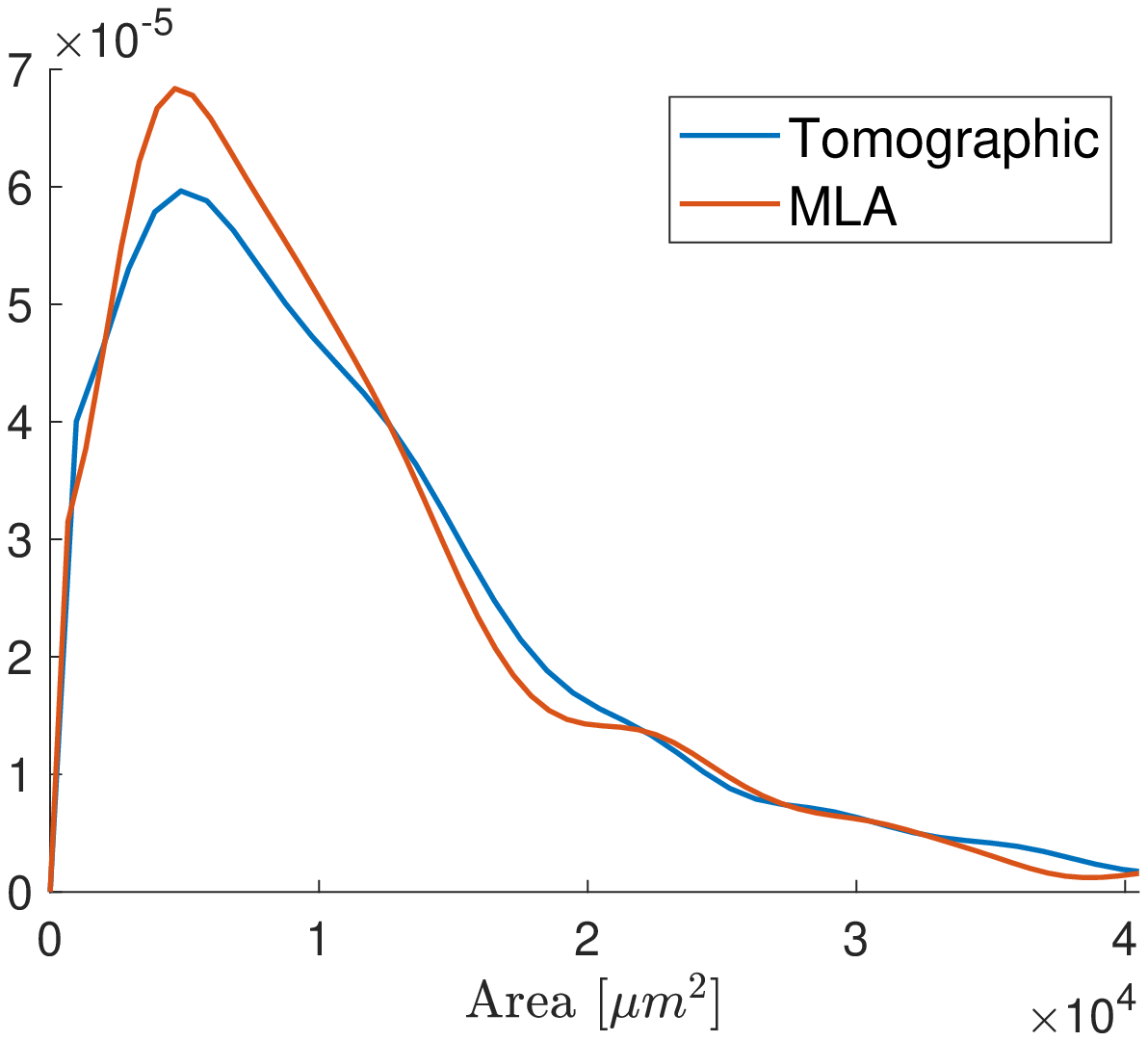} 
\includegraphics[width=0.3\textwidth]{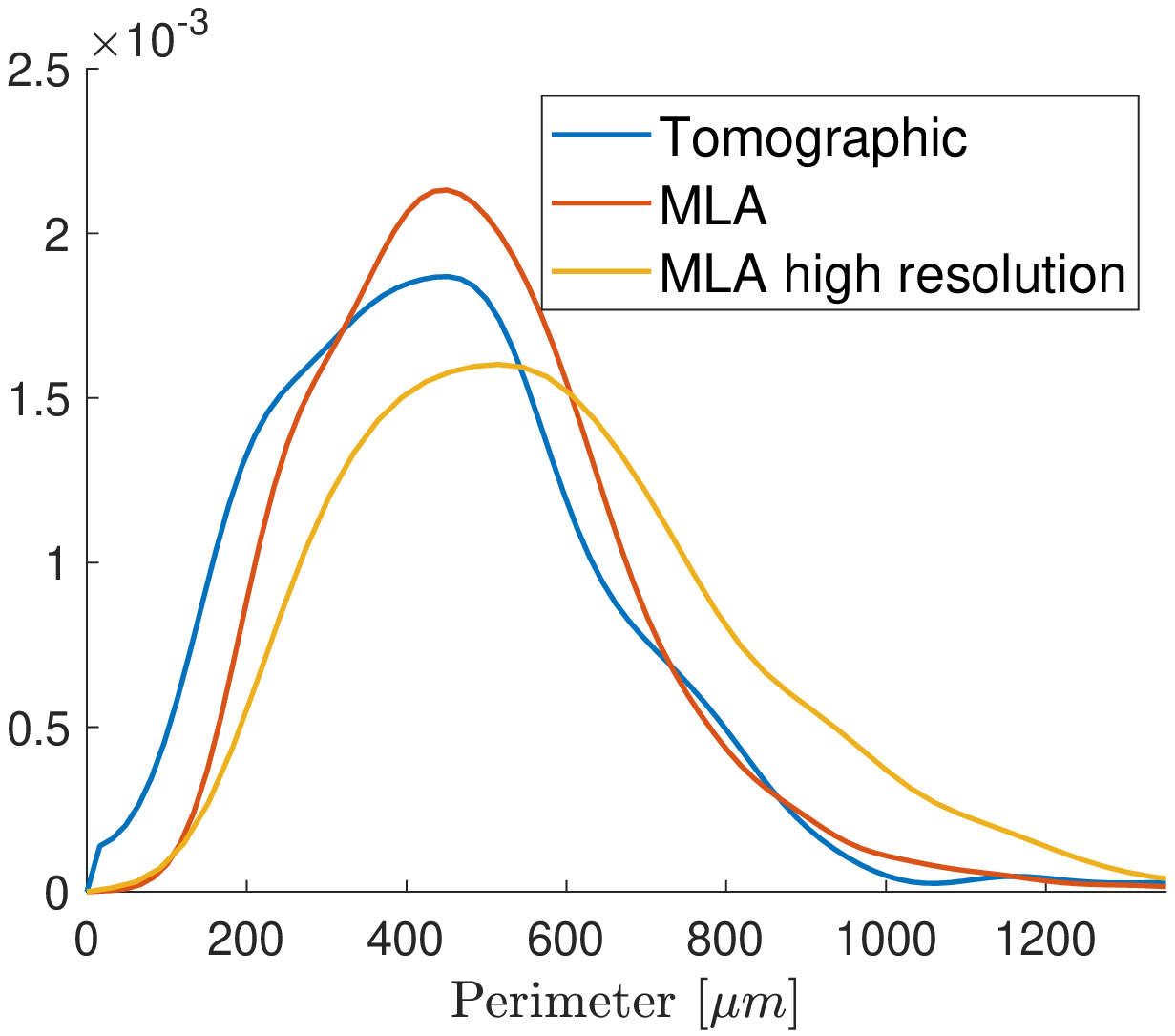}
\includegraphics[width=0.3\textwidth]{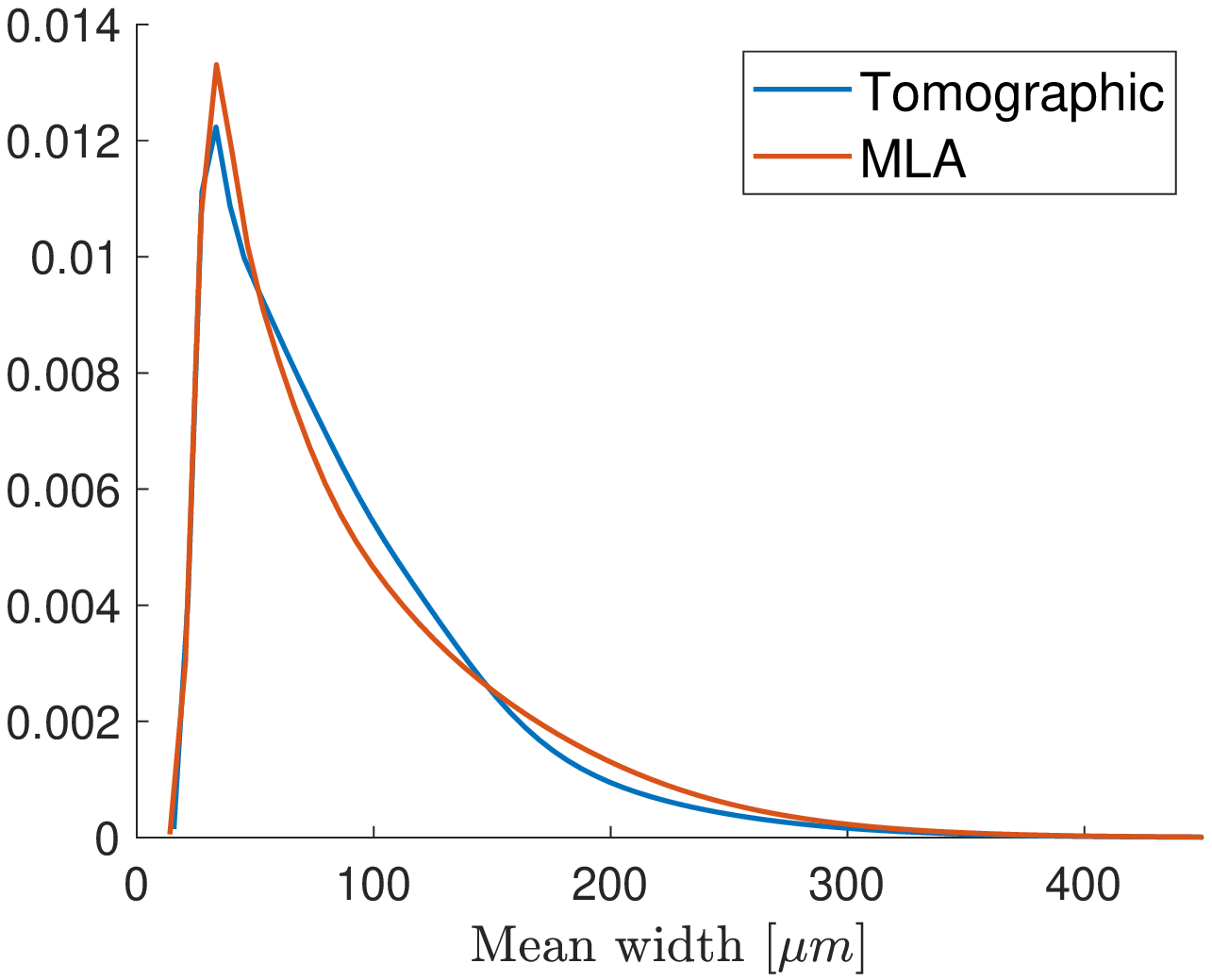}\\
\begin{minipage}{0.3\textwidth}{\centerline{a)}}\end{minipage}
\begin{minipage}{0.3\textwidth}{\centerline{b)}}\end{minipage}
\begin{minipage}{0.3\textwidth}{\centerline{c)}}\end{minipage}
\caption{Comparison of distributions of particle-wise size characteristics for MLA data and two-dimensional cross-sections of the 3D data. a) Distribution of particle area, based on MLA data and cross-sections of the segmented 3D particle system. b) Distribution of particle perimeter length. The MLA data was coarsened in order to get more comparable results. c) Distribution of the mean width of particles.}
\label{fig:area}
\end{center}
\end{figure}

 In Figure \ref{fig:area}, one can observe that the distributions of these size characteristics are consistent for the MLA data set and the planar cross sections of the 3D data. Since the perimeter of particles is resolution sensitive and because the MLA images have a much higher resolution than the XMT image, the perimeters are generally larger in the MLA case, see Figure \ref{fig:area} b). For better comparability we therefore coarsened the particles in the MLA data such that we have the same resolution as in the 3D case. Nevertheless the particle perimeters obtained by the coarsened MLA data are still a bit larger, which can be explained by the morphological smoothing during the segmentation process of the XMT image.

\subsubsection{Shape characteristics}
Various shape characteristics can be deduced for two-dimensional particle cross-sections, see \cite{chiu2013}. Among these characteristics, we use the sphericity factor
\begin{equation}
s(P)=4 \pi \frac{a(P)}{l(P)^2}.
\end{equation}
The sphericity factor takes values between $0$ and $1$, where the value $1$ is achieved for a circular particle and lower values indicate higher deviation from the shape of a circle.

Another shape characteristic is the convexity factor, which is defined by
\begin{equation}
c(P)=\frac{a(P)}{a{\left(q(P)\right)}},
\end{equation} 
where $q(P)$ is the convex hull of $P$. In this case, the value $1$ is obtained for convex particles and lower values indicate higher deviations from convexity.

Finally, we consider the elongation factor defined by 
\begin{equation}
e(P)=\frac{\ell_1 {\left(\varepsilon(P)\right)}}{\ell_2 {\left(\varepsilon(P)\right)}},
\end{equation}
where $\varepsilon(P)$ is the best fitting ellipsoid to $P$, $\ell_1 {\left(\varepsilon(P)\right)}$ length of its short semiaxis, and $\ell_2 {\left(\varepsilon(P)\right)}$ length of its long semiaxis. Similarly to the sphericity factor, the value 1 is, among others, achieved for a circle. However, this quantity does not depend on the surface area and is less sensitive to irregularities of the particle surface. Note that the best fitting ellipsoid can be found by principal component analysis \citep{macsleyne2008}.

\begin{figure}
\begin{center}
\includegraphics[width=0.3\textwidth]{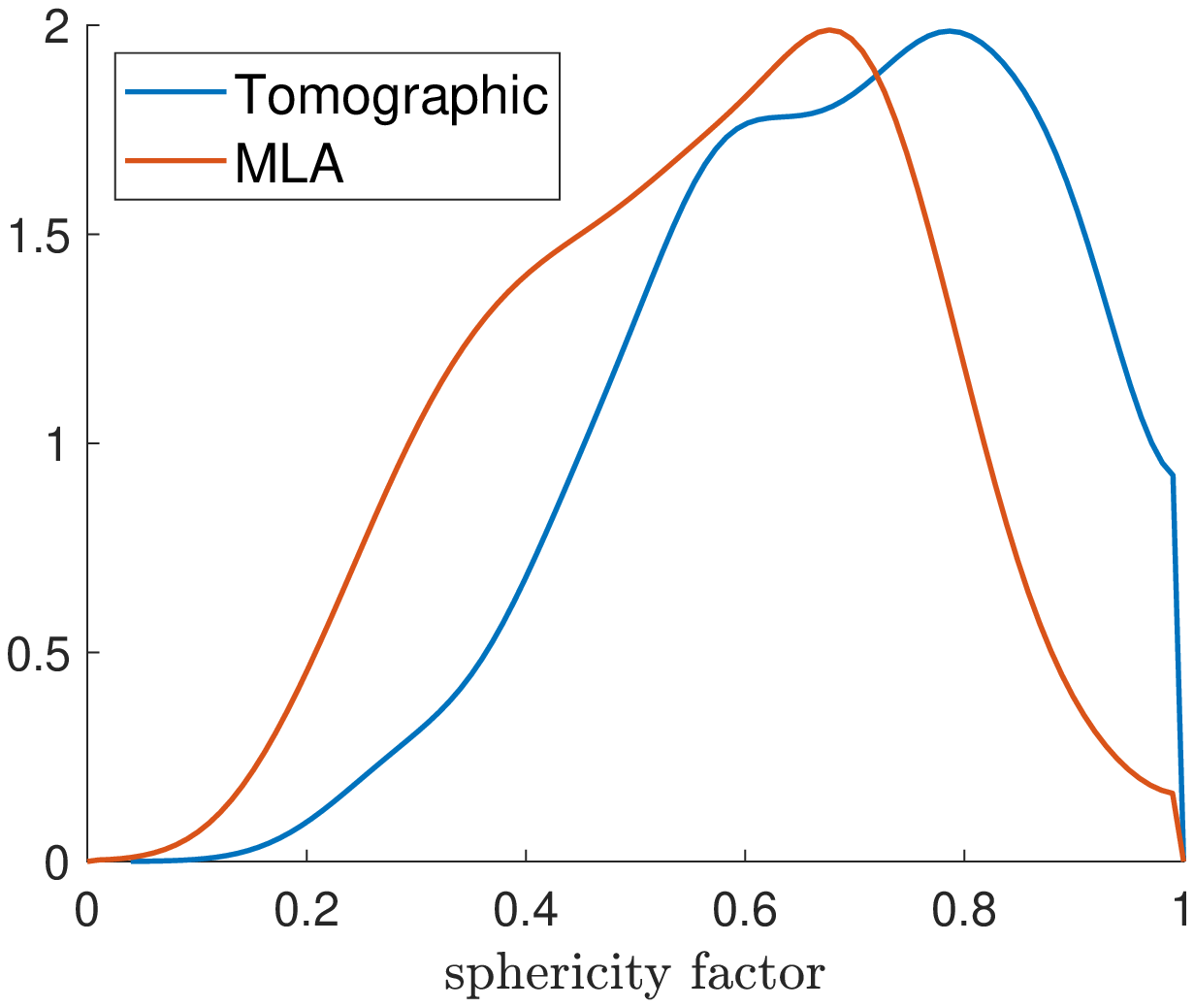} 
\includegraphics[width=0.3\textwidth]{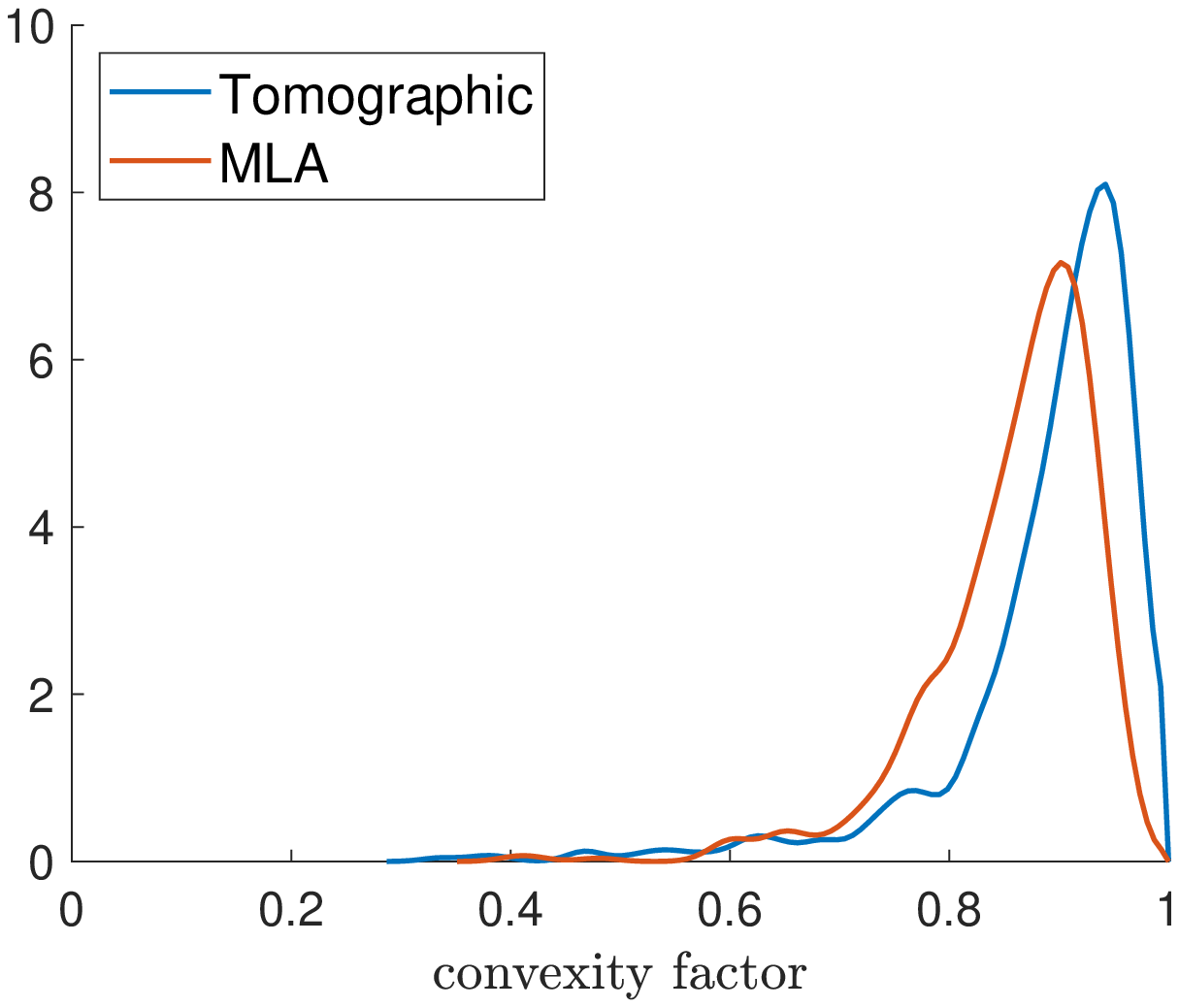}
\includegraphics[width=0.3\textwidth]{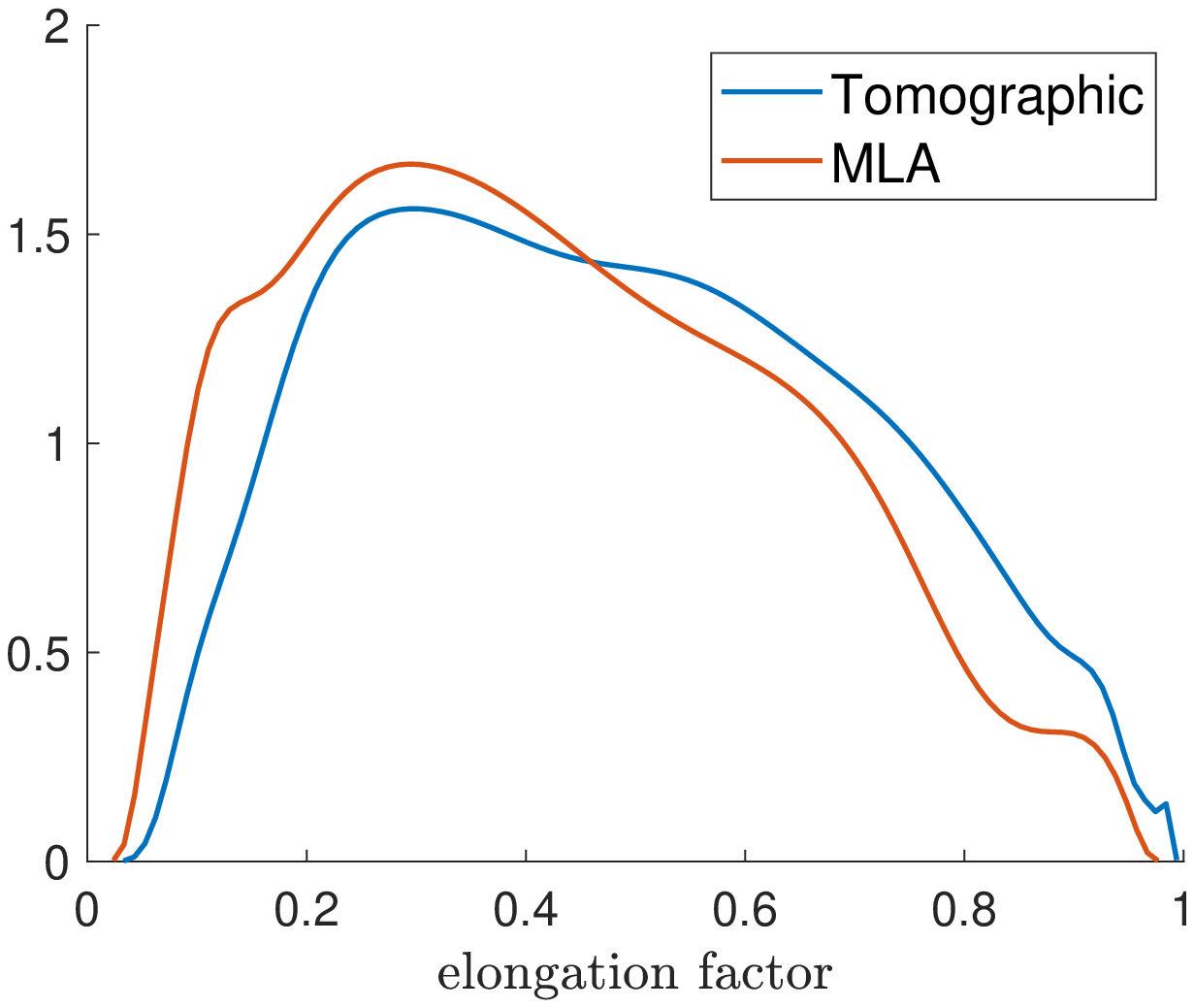}\\
\begin{minipage}{0.3\textwidth}{\centerline{a)}}\end{minipage}
\begin{minipage}{0.3\textwidth}{\centerline{b)}}\end{minipage}
\begin{minipage}{0.3\textwidth}{\centerline{c)}}\end{minipage}
\caption{Comparison of distributions of particle-wise shape characteristics for MLA data and two-dimensional cross-sections of the 3D data. a) Distribution of sphericity factor. b)  Distribution of convexity factor. c) Distribution of elongation factor. }
\label{fig:shape}
\end{center}
\end{figure}
A comparison between the distributions of shape characteristics for MLA data and tomographic data is shown in Figure \ref{fig:shape}. The considered shape characteristics seem to be slightly larger for the tomographic data. This can also be explained by the morphological smoothing during the segmentation process explained in Sections \ref{sec:imageProcessing} and \ref{sec:segmentation}, which produces in general more spherical and convex shapes.
Nevertheless both the shape and size characteristics are quite consistent for the two data sets. In the next section we will describe how to find the location of a 2D MLA image in the 3D data and give an example of how to use the additional information which MLA images provide.

\subsection{Registration and prediction of attenuation coefficients}
The grayscale value $I(x)$ of a voxel $x$ in the 3D image is closely related to the local X-ray attenuation coefficient of the material at location $x$. Specifically, there is a well-known monotone relationship between the grayscale value $I(x)$ and the value of the product $\rho(x) \mu_m(x)$, where $\rho(x)$ is the mass density of the material at location $x$ and $\mu_m(x)$ its mass attenuation coefficient, see \cite{pavlinsky2008}. Still, the grayscale values of voxels in the 3D image describe the value of $\rho(x) \mu_m(x)$ only qualitatively, meaning that brighter voxels indicate higher X-ray absorption which, in conclusion, indicates larger volumetric mass densities or mass attenuation coefficients. In this section we will assume a linear relationship $\rho(x) \mu_m(x)=a I(x)+b$ where the regression parameters $a,b\in \R$ are unknown. To find a quantitative relationship, meaning determining the constants $a$ and $b$, between the grayscale values $I(x)$ in the 3D data and the values of $\rho(x) \mu_m(x)$ we can use information from the MLA data. To be more precise, we first register a 2D MLA image which can be correlated with a planar section of the volumetric 3D data. This will allow us to compare the grayscale values of the 3D data with the mass density and mass attenuation coefficient of the corresponding mineral observed in the MLA image.
 \subsubsection{Registration}\label{subsec:Registration}
We now describe in detail how we located the 2D MLA images in the 3D image. Let $B_{\text{MLA}}, B$ be binarized 2D and 3D images, respectively. The binarization of the 3D XMT image was described in detail in Section~\ref{sec:binarization} and since the 2D MLA images are false color images, where each color either represents a mineral phase or the background, it is easy to binarize such an image. 

\begin{figure}
 \begin{center}
 \includegraphics[width=0.45\textwidth]{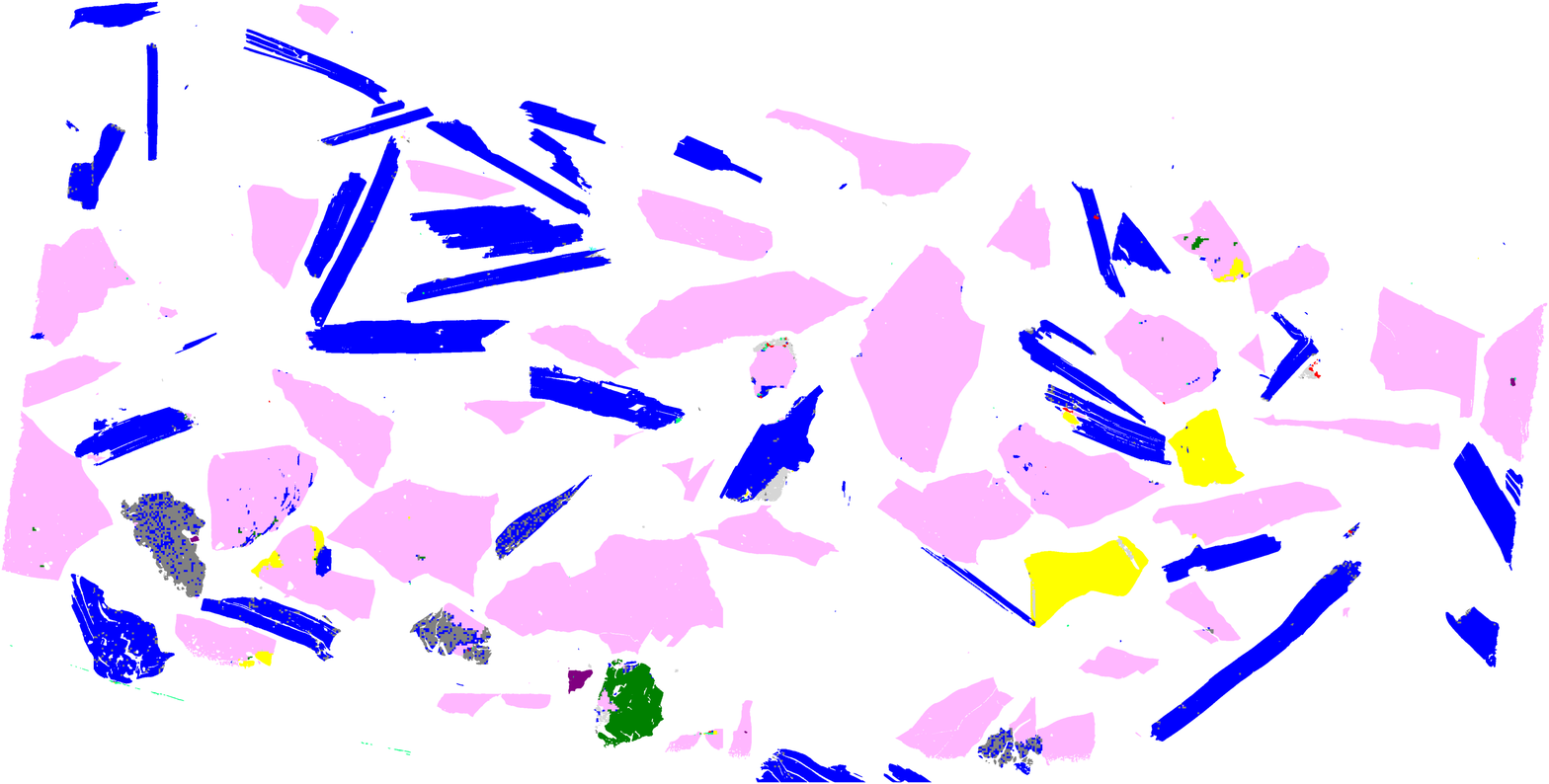} 
 \includegraphics[width=0.45\textwidth]{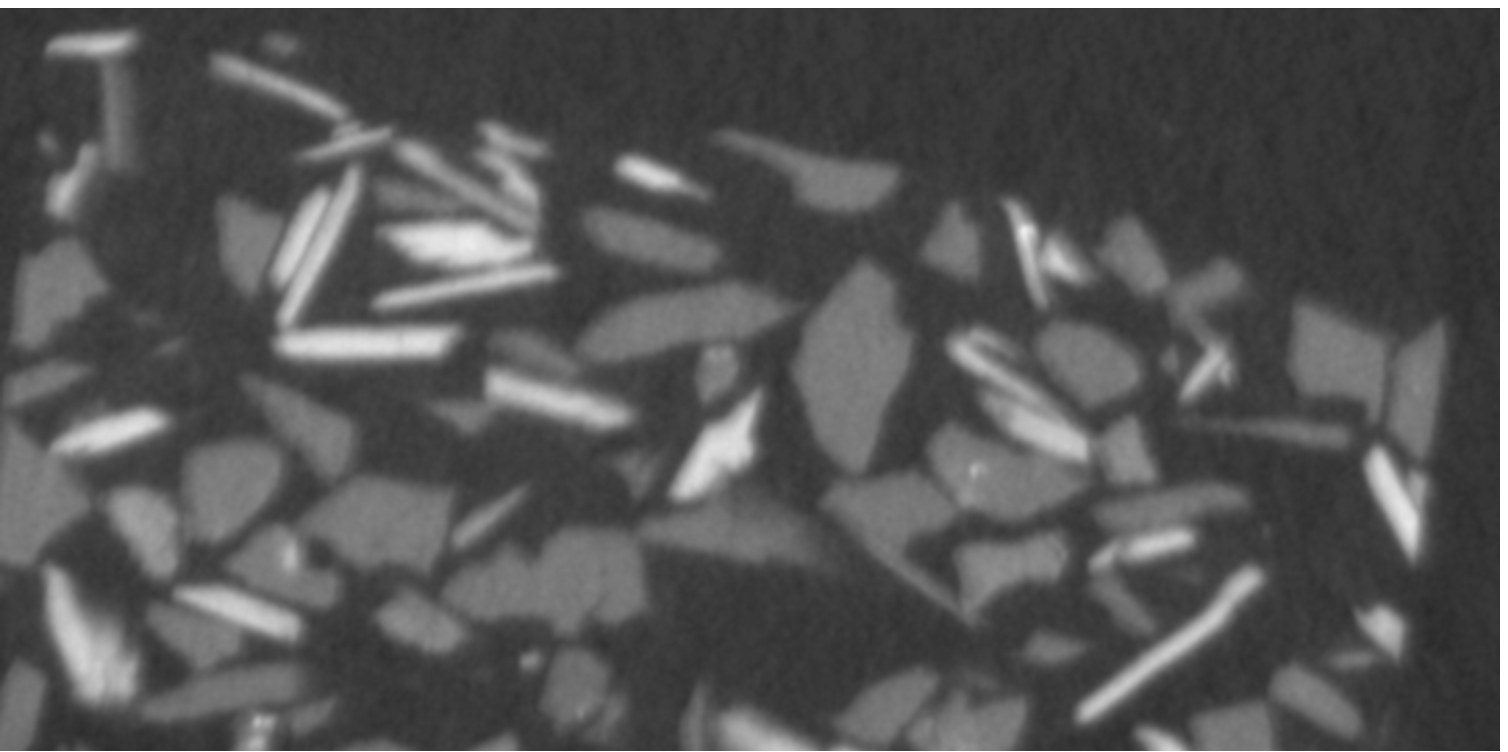}\\
 \begin{minipage}{0.45\textwidth}{\centerline{a)}}\end{minipage}
 \begin{minipage}{0.45\textwidth}{\centerline{b)}}\end{minipage}
 \caption{Registration of the 2D MLA image within the 3D image. a) 2D MLA image. The colors indicate different minerals, e.g. blue indicates Zinnwaldite. b) The corresponding planar section in the 3D XMT image localized by means of (\ref{eq:registration}).}\label{fig:registration}
 \end{center}
 \end{figure}

For a rotation matrix $R \in SO_3$ we denote the correspondingly rotated 3D image by $B_R$. The location of the 2D image in the 3D image is described by the rigid transformation consisting of some $R_0 \in SO_3$, describing the rotation of the 3D image, and $x_0\in \R^3$ which is the shift of the 2D image. That is
\begin{equation}\label{eq:registration}
(x_0,R_0)=\argmax_{(x,R) \in \R^3 \times SO_3} \sum\limits_{y\in \Z^3} B_R (y) B_{\text{MLA}}(y-x),
\end{equation}
where the image values outside of their corresponding observation windows are set equal to $0$. In particular, for the 2D image we have $B_\text{MLA}(x_1,x_2,x_3)=0$ if $x_3 \neq 0.$ The optimization problem described in (\ref{eq:registration}) was solved with the Nelder-Mead method, see \cite{nelder1965}. By expressing the sum on the right-hand side of (\ref{eq:registration}) as a convolution, i.e.,
\begin{equation}
\sum\limits_{y\in \Z^3} B_R (y) B_{\text{MLA}}(y-x)=(B_R \ast \widetilde{B}_\text{MLA}) (x),
\end{equation}
where $\widetilde{B}_\text{MLA}(x)=B_\text{MLA}(-x)$ for an arbitrary rotation matrix $R \in SO_3$, we can use the fast Fourier transformation to accelerate computations, see for example \cite{Burger2010}. Further acceleration can be achieved by upscaling the images to determine a good start configuration $(x,R)\in \R^3 \times SO_3$ for the optimization at the given scale. Results of the registration process are visualized in Figure \ref{fig:registration}. 

 \subsubsection{Prediction of attenuation coefficients}
  Due to the registration of the MLA images described in Section \ref{subsec:Registration} we now have information about the mineralogical composition for some planar sections of the 3D particle system. We show how this information can be used for predicting local material specific constants based on grayscale values.  In this section we will use only one of the given MLA images for the calibration of the prediction model, and the other MLA image for validation.
  
  For that purpose let $I \colon W^\prime \to \{0,\dots, 65535\}$ be the 3D grayscale image and $I_\text{MLA}\colon V^\prime \to \{0,\dots, 255\}$ be a registered MLA image, meaning $I_\text{MLA}$ is a 2D MLA image after the rigid transformation given by (\ref{eq:registration}) with $V^\prime \subset W^\prime.$ It is important to note that, contrary to the tomographic image $I$, the value $I_\text{MLA}(x)$ indicates which mineral is present at the location $x \in V^\prime$. For example, in our data $I_\text{MLA}(x)=19$ means that quartz was observed at $x$, allowing us to compute the quartz phase 
  \begin{equation}\label{eq:quartzPhase}
\Omega_\text{Quartz} = \{x \in V^\prime : I_\text{MLA}(x)=19 \} \subset V^\prime
  \end{equation}
  in the planar section $V^\prime.$ This information about the quartz phase, gained from the MLA image, can be transfered to the 3D tomographic image. To be precise, using (\ref{eq:quartzPhase}), we can compute the mean grayscale value of voxels associated with quartz in the 3D image by
 \begin{equation}\label{eq:meangrayscale}
 \bar{I}_\text{Quartz}=\frac{1}{|\Omega_\text{Quartz}|} \sum\limits_{x\in \Omega_\text{Quartz}} I(x),
 \end{equation}
 where $|\Omega_\text{Quartz}|$ denotes the number of voxels in $\Omega_\text{Quartz}.$
 Analogously, the mean grayscale value can be computed for other minerals depicted in the MLA image.
  \begin{table}
  	\centering
  	{\renewcommand{\arraystretch}{1.5}
  \input{table3.tex}
}
      \caption{Minerals observed in the MLA image and their corresponding mass density $\rho$, mass attenuation coefficient $\mu_m$ and mean grayscale value $\bar{I}$ in the 3D XMT image.}\label{table:grayscale}
  \end{table}
 Table \ref{table:grayscale} lists some minerals, which can be observed in a sufficient quantity in the MLA image, and their corresponding mean grayscale values, which were determined by means of (\ref{eq:meangrayscale}). Furthermore, for each of these minerals the mass density $\rho$ can be found in \cite{handbookOfMineralogy} and there are lists of the mass attenuation coefficient $\mu_m$ of many elements from the periodic table, which can be used for the estimation of $\mu_m$ for chemical compounds like minerals, see \cite{hubbell1995}. This allows us to correlate the mean grayscale value of each mineral type with their $\rho$ and $\mu_m$ values.
  
\begin{figure}[!htb]
  \begin{center}
  \includegraphics[width=0.45\textwidth]{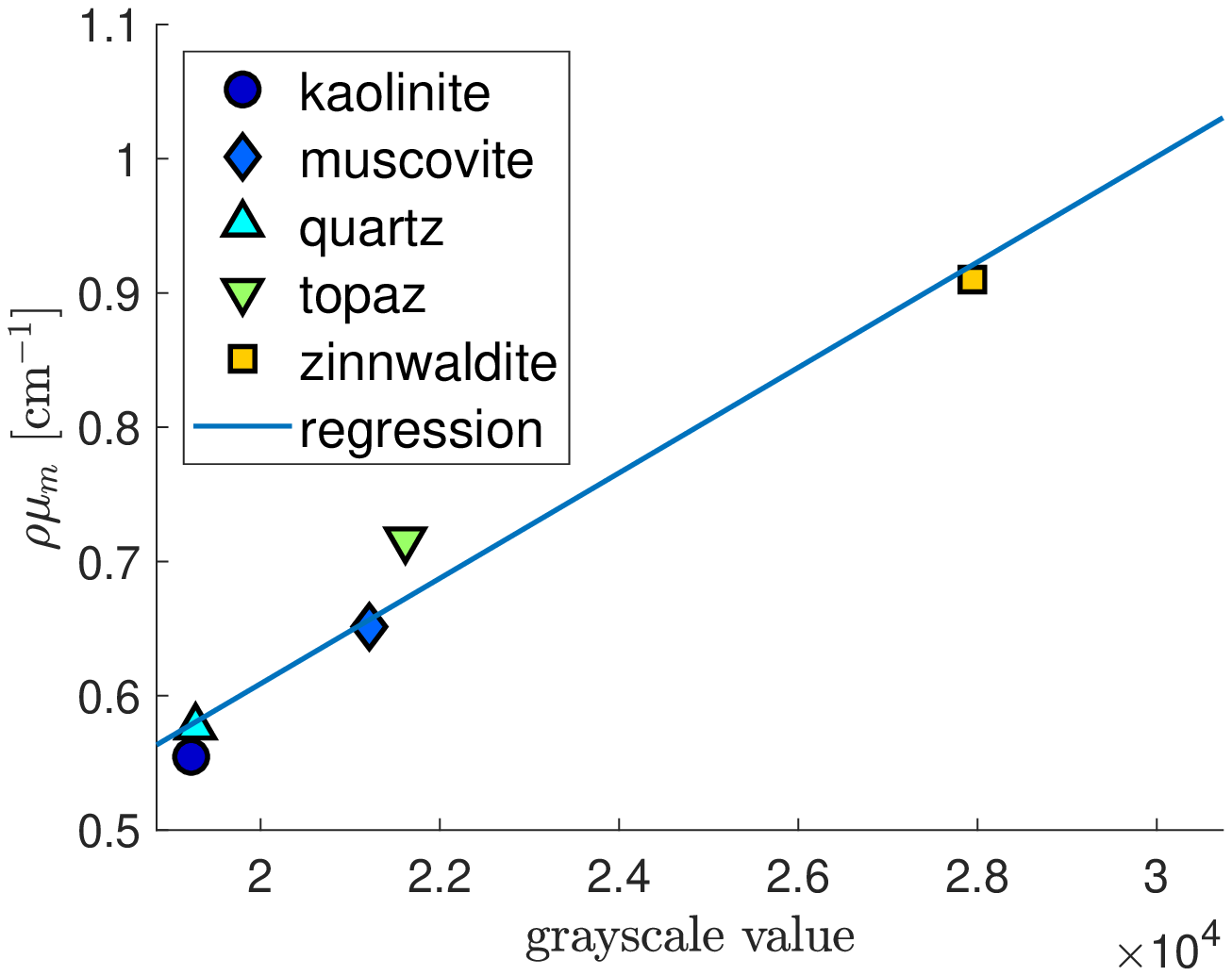} 
  \includegraphics[width=0.45\textwidth]{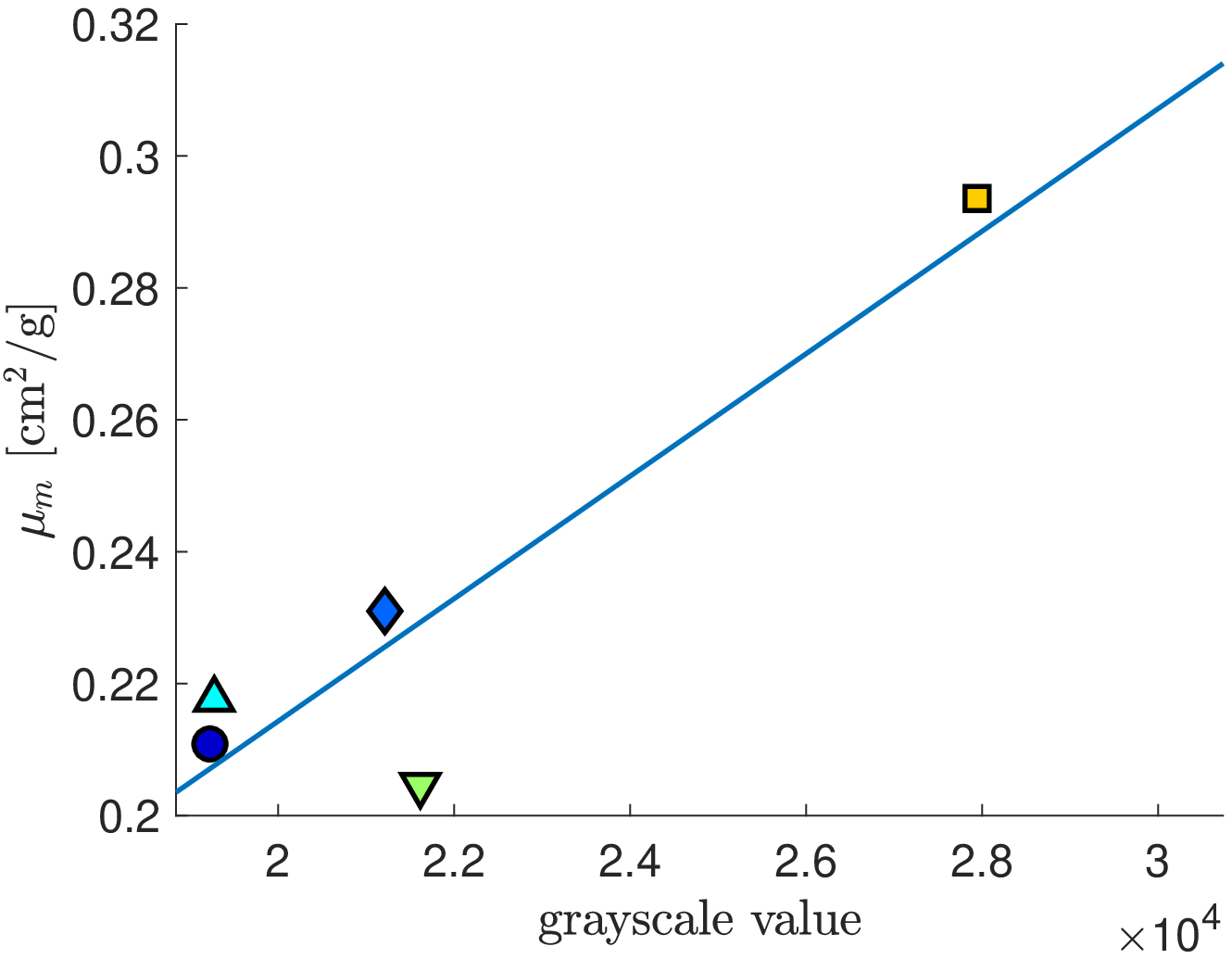} 
  \begin{minipage}{0.45\textwidth}{\centerline{a)}}\end{minipage}
  \begin{minipage}{0.45\textwidth}{\centerline{b)}}\end{minipage}
  \end{center}
  \caption{Mean grayscale values of different minerals in the 3D XMT image plotted against their material specific properties. The association of voxels in the XMT image with the different minerals was established by registration of the MLA image. a) Linear relationship between the mean grayscale values in the XMT image and the $\rho \mu_m$ values. b) Relationship between the mean grayscale values and the mass attenuation coefficient $\mu_m$ under the assumption of a constant mass density $\rho$.}
  \label{fig:grayscale_vs_atomic}
  \end{figure}

For example, $|\Omega_\text{Quartz}|=203980$ voxels in the 3D image were detected which represent quartz according to the MLA image. These voxels have a mean grayscale value of $ \bar{I}_\text{Quartz}=19274$ in the 3D image. Zinnwaldite, on the other hand, which has a higher mass density and mass attenuation coefficient, has a mean grayscale value of $ \bar{I}_\text{Zinnwaldite}=27943$. Figure \ref{fig:grayscale_vs_atomic} a) visualizes the linear relationship between mean grayscale values of several minerals and their $\rho \mu_m$ values. By means of linear regression, see \cite{hastie2009}, with the data from Table \ref{table:grayscale} we obtain the relationship
  \begin{equation}\label{eq:regression1}
  \rho \mu_m =3.9 \cdot 10^{-5} I-0.17,
  \end{equation}
  for grayscale values $I\in \{0,\dots,65535\}.$ One should note that (\ref{eq:regression1}) is not suitable for extrapolation, especially since for grayscale values smaller than $4500$ the estimated material constants become negative.
  Nevertheless, (\ref{eq:regression1}) estimates the local $\rho \mu_m$ values very well for grayscale values between $1.8 \cdot 10^4$ and $3 \cdot 10^4$. 
 
\begin{figure}[!htb]
  \begin{center}
  \includegraphics[width=0.45\textwidth]{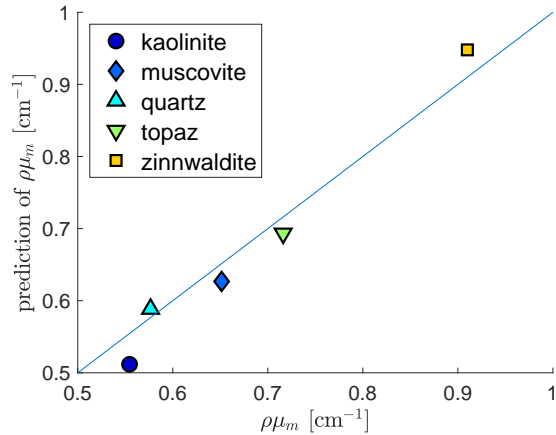} 
  \end{center}
  \caption{The predicted $\rho \mu_m$ value derived from mean grayscale values of different minerals observed in a planar section which was not used for the calibration of the regression (\ref{eq:regression1}) versus the actual $\rho \mu_m$ values given by MLA. Points near the blue line $x=y$ indicate a good prediction.}
  \label{fig:validation}
  \end{figure}

 Since only one of the two MLA images was taken into account for the calibration of the regression given by (\ref{eq:regression1}) the other MLA image can be used for validation.
 Therefore, analogously to (\ref{eq:quartzPhase})--(\ref{eq:meangrayscale}), we determined the mineral phases along a second planar section provided by the second MLA measurement and determined the mean grayscale values for these phases. With the help of the prediction formula (\ref{eq:regression1}) we can then estimate the $\rho \mu_m$ values based on the grayscale values and compare these with the true $\rho \mu_m$ values given by the underlying MLA measurement.
Figure~\ref{fig:validation} shows that Formula (\ref{eq:regression1}) works rather well for predicting local $\rho \mu_m$ values in areas of the image which were not used for the calibration of the regression model. This means that, with the help of MLA images, we are able to determine material specific quantities solely based on grayscale values, thus allowing us to some degree the determination of the mineralogical composition in the entire 3D XMT image. For example, zinnwaldite has a rather unique $\rho \mu_m$ value among the minerals of the sample which allows us to distinguish it from other minerals based on its grayscale value.  
  
  Still, it is not always possible to determine the underlying mineral solely based on its $\rho \mu_m$ value, because, for example, quartz and kaolinite have similar  $\rho \mu_m$ values. 
  A direct relationship between the mean grayscale value and the mass attenuation coefficient $\mu_m$ can be established under the assumption of constant mass densities of the minerals. 
  For that purpose (\ref{eq:regression1}) suggests a regression of the type
    \begin{equation}
   \mu_m=c_1 I +c_2,
    \end{equation}
 where the parameters $c_1,c_2 \in \R$ have to be determined. However, Figure~\ref{fig:grayscale_vs_atomic} shows that this approach does not work due to the different mass densities among the minerals of our sample. Therefore, although we can differentiate some minerals by means of (\ref{eq:regression1}) based on their $\rho \mu_m$ values, which are determined by their grayscale values, some uncertainty remains due to the fact that $\rho \mu_m$ values are not unique to certain minerals.

\section{Conclusions}
We presented a method to segment 3D XMT image data of particle systems by combining the marker-based watershed algorithm with a post processing step which utilizes neural networks. 
By reducing oversegmentation -- a common issue of the watershed algorithm -- this led to a system of properly segmented 3D particles. We then compared the 3D segmentation with high resolution segmentations of planar sections obtained by MLA, by comparing distributions of several size and shape characteristics. Furthermore, we described a procedure to embed 2D MLA images into the 3D sample using rigid transformations. This gave us additional information about the mineralogical composition of the 3D XMT image along planar sections. Thus we were able to find a quantitative relationship between grayscale values in the 3D image and material specific constants, like the mass density $\rho$ and the mass attenuation coefficient $\mu_m$. In a forthcoming study we will extend this approach by analyzing material-specific shape and size characteristics based on the segmentation presented in Section \ref{sec:segmentation} and combining these characteristics with grayscale information from XMT data. We expect that such an approach can then allow the prediction of the mineralogical composition of a particle, when MLA information is not available and grayscale values alone are not sufficient for making such decisions.

\section{Acknowledgements}
The financial support of the German Research Foundation (DFG) for funding the X-ray microscope (INST267/129-1) as well as the research projects (PE1160/22-1 and SCHM997/27-1) within the priority program SPP 2045 ``Highly specific and multidimensional fractionation of fine particle systems with technical relevance'' is gratefully acknowledged.
The authors would like to thank Sabine Gilbricht for her work at the MLA and Roland W\"urkert for preparing the epoxy blocks.

\newpage
\bibliography{Bibliography}

\end{document}

%% file: table1.tex
	\begin{tabular} {l l|c}
			\bf{parameter} & & \bf{value} \\ \hline
            source position & mm & -35 \\
            detector position & mm & 18 \\
            objective & & 4X \\
			camera binning & & 2 \\
            magnification & & 6.06 \\
			pixel size & \textmu m & 4.5 \\
            voltage/power & kV/W & 70/5 \\
            filter & & LE3 \\
            exposure time & s & 8 \\
            angle & grad & 360 \\
            projections & & 2401 \\
            scan time & h:mm & 9:30 \\            
        \end{tabular}

%% file: table2.tex
	\begin{tabular} {l|c}
			\bf{function} & \bf{parameter}\\ \hline
            center shift & automatic \\
            smoothing & Gaussian, 0.7 \\
			beam hardening constant & 0 \\
			byte scaling & manual (-200; 1500) \\
            defect correction & bright and dark spots \\
         \end{tabular}

%% file: table3.tex
  \begin{tabular}{c|*5c}\hline
  \bfseries mineral & quartz & kaolinite & muscovite & zinnwaldite & topaz \\\hline
 $\bm{\rho}$ [g/cm$^3$] & 2.65 & 2.63 & 2.82 & 3.1 & 3.5\\\hline
  $\bm{\mu_m}$ [cm$^2$/g]& 0.22 & 0.21 & 0.23 & 0.29 & 0.2 \\\hline
  $\bm{\bar{I}}$ & 19274 & 19225 & 21213 & 27943 & 21615 \\\hline
  \end{tabular}